\documentclass[12pt]{article}%
\usepackage{amsmath}
\usepackage{amsfonts}
\usepackage{amssymb}
\usepackage{graphicx}%
\providecommand{\U}[1]{\protect\rule{.1in}{.1in}}
\begin{document}

\title{Measurement of the Geometrical Decay of the Spin Hall Effect in Fe(CsAu)$_{\nu
}$ Multi Layers}
\author{F.Song, H.Beckmann and G.Bergmann\\Department of Physics\\University of Southern California\\Los Angeles, California 90089-0484\\e-mail: bergmann@usc.edu}
\date{\today }
\maketitle

\begin{abstract}
The anomalous Hall effect of Fe(CsAu)$_{\nu}$ is investigated ($\nu$ is an
integer). Electrons with spin up and down experience a different degree of
specular reflection at the FeCs interface. This yields a different mean free
path for the two spin orientations. In the presence of an electric field
parallel to the film plane one obtains a spin current in addition to the
charge current. If one introduces impurities with a large spin-orbit
scattering into the Cs host then the combination of spin current and
spin-orbit scattering yields an anomalous Hall effect. By building in situ
multi layers of CsAu ($5nm$ of Cs and 0.04 atomic layers of Au) on top of an
Fe film one can measure the relative magnitude of the spin current normal to
the film. Within the accuracy of the experimental data the spin current in
Fe(CsAu)$_{\nu}$ decays exponentially with a decay length of $20nm$.


\newpage

\end{abstract}

\section{Introduction}

The investigation of spin currents and the anomalous Hall effect (AHE) have
been attracting considerable interest in the past decades. The possible
applications in spintronics \cite{S53}, \cite{B141}, \cite{A61}, \cite{A60}
will further accelerate the exploration of this interesting field.

An impurity with a strong spin-orbit interaction scatters spin up and down
electrons with opposite left-right asymmetry. The asymmetric scattering of
conduction electrons by spin-orbit scatterers was already investigated by
Ballentine and Huberman \cite{B62}, \cite{H17} almost 30 years ago. They
explained the deviations of the Hall constant from the free electron values
for heavy liquid metals such as Tl, Pb and Bi. They calculated the spin-orbit
scattering using perturbation theory. One of the authors \cite{B126} has
calculated exactly the AHE of a polarized electron gas in the presence of
spin-orbit scattering (SOS) in terms of Friedel phase shifts.

The opposite left-right asymmetry of spin up and down scattering by spin-orbit
interaction yields two related effect in zero magnetic field:

\begin{itemize}
\item In the presence of a spin current SOS yields an anomalous Hall effect. A
spin current through a conductor with spin-orbit scattering yields a net
transverse scattering and therefore a net transverse electric field. The spin
up and down electrons in a spin current have opposite (drift-) velocities and
therefore a spin-orbit scatterer deflects both spins to the same side. This
generates an anomalous Hall effect, i.e., spin Hall effect with a transverse
electric field.

\item For a pure charge current the spin up and down electrons are scattered
to opposite sides of the original trajectory. The SOS does not generate a
charge imbalance. However, it creates opposite gradients in the chemical
potential for spin up and down electrons. This effect has been predicted by
Dyakonov and Perel \cite{D42} and Hirsch \cite{H24}. It has been recently
observed in semiconducting samples \cite{A70}, \cite{W37}.
\end{itemize}

Both phenomena are two sides of the same effect. This effect became popular in
recent years under the name of "Spin Hall Effect".

In this paper we investigate spin currents with the help of spin-orbit
scattering impurities. Recently our group investigated the anomalous Hall
effect (AHE) in double layers of FeCs which were covered with sub-mono layers
of Pb and Au \cite{B137}. When the double layer FeCs with $d_{Fe}=12.4nm$ and
$d_{Cs}=29.6nm$ was covered with $1/50$ of a mono layers of Pb it showed a
remarkable result. The AHE increased by about a factor of five when it was
covered with $1/50$ of a mono layers of Pb. (The geometry of the layers is
shown in Fig.1a). In this spatial configuration the contribution of a Pb atom
to the AHE was stronger by a factor of $2\times10^{4}\ $than that of an Fe
atom. If the FeCs double layer was covered with $1/50$ atomic layer of Au then
the AHE had the opposite sign and was about a factor four smaller.%

\begin{align*}
&
%TCIMACRO{\FRAME{itbpF}{5.335in}{2.7015in}{0in}{}{}{fecsau.eps}%
%{\special{ language "Scientific Word";  type "GRAPHIC";
%maintain-aspect-ratio TRUE;  display "USEDEF";  valid_file "F";
%width 5.335in;  height 2.7015in;  depth 0in;  original-width 5.0668in;
%original-height 2.5521in;  cropleft "0";  croptop "1";  cropright "1";
%cropbottom "0";  filename 'FeCsAu.eps';file-properties "XNPEU";}}}%
%BeginExpansion
{\includegraphics[
height=2.7015in,
width=5.335in
]%
{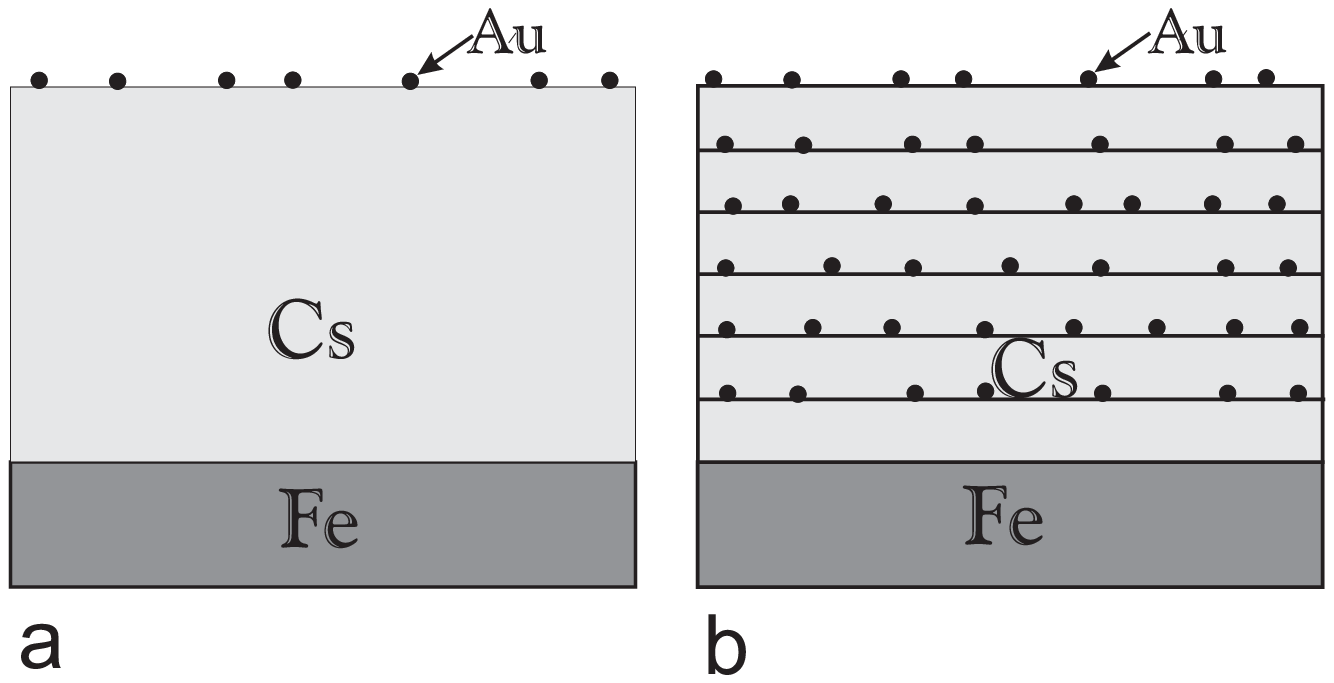}%
}%
%EndExpansion
\\
&
\begin{tabular}
[c]{l}%
Fig.1a: A single Au coverage of an FeCs double layer.\\
Fig.1b: The geometry of Fe(CsAu)$_{\nu}$ multi-layers. The thickness of the Fe
film is\\
$d_{Fe}=9.0nm,$ the repeated multi-layers of \ CsAu have $d_{Cs}=5nm$ and a
Au\\
coverage of 0.04 atomic layer.
\end{tabular}
\end{align*}

Since Pb and Au are non-magnetic they don't yield a magnetic AHE. Instead the
AHE has to be due to a spin current in the Cs. The origin of the spin current
is due to the fact that spin up and down electrons experience a different
exchange potential on the Fe side of the FeCs interface. This yields a
different specular reflection of spin up and down electrons at the interface.
As a consequence spin up and down electrons have different mean free paths
(MFP) and, in the presence of an electric field, different drift velocities.
(The large MFP in the Cs and the high degree of specular reflection at the
upper surface (of the order of 80-90\%) enhanced this effect). The spin
current is the difference between the spin up and down currents. Since the
host Cs is non magnetic the carrier concentration of spin up and down
electrons is equal to half the electron density $n/2$ and the spin current is
equal to $j_{s}=n\left(  -e\right)  \left(  v_{+}-v_{-}\right)  /2$. Our
observation of the AHE in FeCsPb layers was an early observation of the spin
Hall effect. The mechanism will be discussed in more detail in the discussion
and the appendix.

In this paper we investigate the development of the spin current as a function
of film thickness. We prepare multi-layers consisting of an Fe film which is
covered in several steps with sequences of $5nm$ of Cs and an impurity
coverage of 0.04 atomic layer of a noble metal (see Fig.1b). For the noble
metal we use (i) the strong spin-orbit scatterer Au and (ii) the weak
spin-orbit scatterer Ag. The Au and Ag impurities differ strongly in their
spin-orbit scattering but are otherwise very similar.

In this paper we use the following abbreviations: AHE=anomalous Hall effect,
AHC=anomalous Hall conductance, AH ..=anomalous Hall .., MFP=mean free path,
SOS=spin-orbit scattering.%
\[
\]

\section{Experiment}

The preparation of the multi-layers and the measurements are performed in situ
in an evaporation cryostat. During the whole experiment the evaporation
cryostat is inserted into the liquid helium bath of a superconducting magnet.
All the walls surrounding the sample are at liquid helium temperature, while
the walls surrounding the evaporation sources are at liquid N$_{2}$
temperature. The vacuum in our system is better than $10^{-11}$torr.

A thin Fe film with a thickness of about $10nm$ and a resistance of about
$100\Omega$ is quench condensed onto a $4.5K$ cold quartz substrate. The Fe
film is annealed to 40 $K$. Then the Fe film is covered with a Cs film of 5
$nm$ and annealed to 12 $K$. In the next step 0.04 atomic layer of Au is
evaporated on top of the Cs film, acting as Au impurities. Again the sandwich
is annealed to 12 $K$. This CsAu evaporation sequence is repeated several
times. The Cs is evaporated from a SAES-Getters evaporation source. The
evaporation rates of the Cs and Au sources are calibrated before and after the
evaporation. After each Au evaporation, we have an Fe(CsAu)$_{\nu}$ sandwich
(where $\nu$ can be 1, 2, 3, ...).

After each evaporation (and annealing) the magneto-resistance and the Hall
resistance of the sample are measured. The measurements are conducted in the
external magnetic field range between -7 and +7 $T$ and at the temperature of
8 $K$. An identical procedure is used to prepare and investigate
Fe(CsAg)$_{\nu}$ multi-layers.

In Fig.2 the conductance $L_{xx}$ of an Fe(CsAu)$_{\nu}$ multi-layer is
plotted as a function of the Cs(Au) thickness $d_{Cs}$. Each time the Cs
surface is covered with 0.04 atomic layer of the noble metal the conductance
is reduced.%

\begin{align*}
&
%TCIMACRO{\FRAME{itbpF}{3.3225in}{2.6218in}{0in}{}{}{lxxcsau.eps}%
%{\special{ language "Scientific Word";  type "GRAPHIC";
%maintain-aspect-ratio TRUE;  display "USEDEF";  valid_file "F";
%width 3.3225in;  height 2.6218in;  depth 0in;  original-width 3.917in;
%original-height 3.0843in;  cropleft "0";  croptop "1";  cropright "1";
%cropbottom "0";  filename 'LxxCsAu.eps';file-properties "XNPEU";}}}%
%BeginExpansion
{\includegraphics[
height=2.6218in,
width=3.3225in
]%
{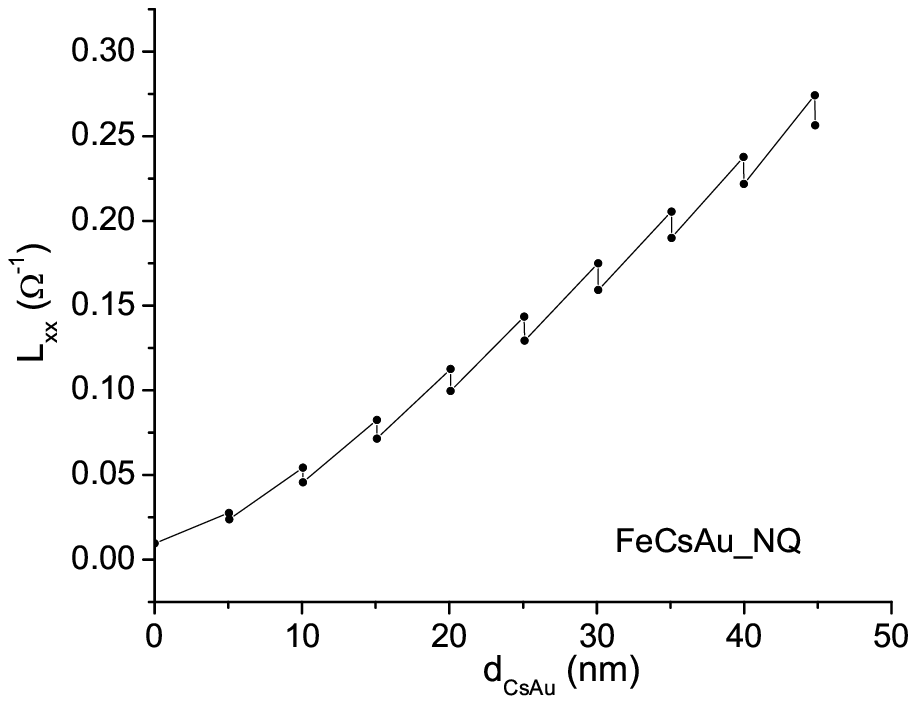}%
}%
%EndExpansion
\\
&
\begin{tabular}
[c]{l}%
Fig.2: The conductance $L_{xx}$ of the Fe(CsAu)$_{\nu}$ multi-layer for\\
successive condensation of the Cs and Au layers. The abscissa\\
is the total thickness of the (CsAu)$_{\nu}$ multi-layers.
\end{tabular}
\end{align*}

In the following we want to compare the AHE between the Fe(CsAu)$_{\nu}$ and
the Fe(CsAg)$_{\nu}$ multi-layers. Therefore it is important that the two
impurities, Au and Ag, have essentially the same scattering cross section for
the host Cs. Besides the large difference in the spin-orbit scattering Au and
Ag are very similar. Both introduce one valence electron and they have an
almost identical atomic volume of $17\times10^{-30}m^{3}.$ We calculated the
effective conductivity of the Cs film $\sigma_{xx}=$ $\left[  L_{xx}\left(
d_{flm}\right)  -Lxx\left(  d_{Fe}\right)  \right]  /\left(  d_{flm}%
-d_{Fe}\right)  $. Here $d_{flm}=d_{Fe}+d_{Cs}$ is the total film thickness.
Using the free electron model $\sigma_{xx}$ yields the effective MFP $l_{eff}$
of the conduction electrons in the Cs with Au impurities. The same calculation
is performed for the Fe(CsAg)$_{\nu}$ multi-layers. In Fig.3, the effective
mean free paths $l_{eff}$ of the conduction electrons in the Cs are plotted
for (CsAu)$_{\nu}$ and CsAg)$_{\nu}$ multi-layers as a function of the total
Cs thickness $d_{Cs}$. The similarity between the Fe(CsAu)$_{\nu}$ and the
Fe(CsAg)$_{\nu}$ multi-layers is quite striking. The first $5nm$ thick Cs film
has an $l_{eff}$ of about $12nm$. Then the coverage with 0.04 atomic layer of
Au or Ag reduces $l_{eff}$ by about $4nm$. The data show that $l_{eff}$
increases with increasing total Cs thickness approaching a value of about
$17nm.$%

\begin{align*}
&
%TCIMACRO{\FRAME{itbpF}{3.149in}{2.567in}{0in}{}{}{mfpcsagau1.eps}%
%{\special{ language "Scientific Word";  type "GRAPHIC";
%maintain-aspect-ratio TRUE;  display "USEDEF";  valid_file "F";
%width 3.149in;  height 2.567in;  depth 0in;  original-width 3.7443in;
%original-height 3.0461in;  cropleft "0";  croptop "1";  cropright "1";
%cropbottom "0";  filename 'MfpCsAgAu1.eps';file-properties "XNPEU";}}}%
%BeginExpansion
{\includegraphics[
height=2.567in,
width=3.149in
]%
{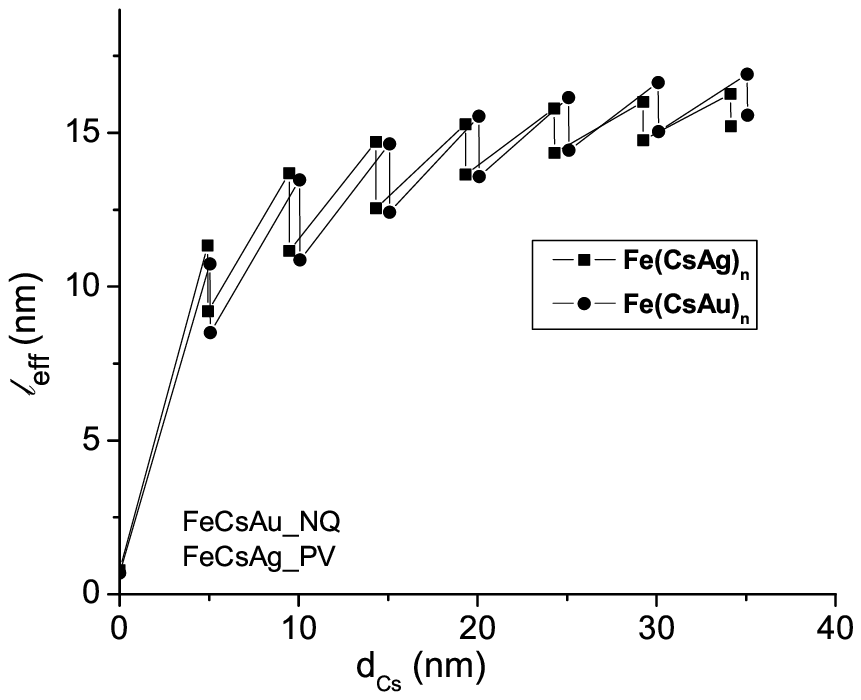}%
}%
%EndExpansion
\\
&
\begin{tabular}
[c]{l}%
Fig.3: The effective mean-free path of the (CsAg)$_{\nu}$ and (CsAu)$_{\nu}$\\
multi-layers as a function of the total Cs film thickness,\\
including the Ag or Au impurities.
\end{tabular}
\end{align*}

The measurement of the Hall resistance yields the AH conductance. In Fig.4 the
anomalous Hall conductance is plotted as a function of the applied magnetic
field for the multi-layers Fe(CsAu)$_{\nu}$ where $1\leq\nu$ $\leq7$. The
normal Hall conductance is subtracted and the anti-symmetric part $\left[
L_{xy}\left(  +B\right)  -L_{xy}(-B)\right]  /2$ is plotted. The AHC is
constant at large fields and can be extrapolated to $B=0$ (as shown in the
upper curve). This yields the AHC $L_{xy,0}$ in zero magnetic field.%

\begin{align*}
&
%TCIMACRO{\FRAME{itbpF}{3.6364in}{2.7854in}{0in}{}{}{lxybcsau.eps}%
%{\special{ language "Scientific Word";  type "GRAPHIC";
%maintain-aspect-ratio TRUE;  display "USEDEF";  valid_file "F";
%width 3.6364in;  height 2.7854in;  depth 0in;  original-width 4.0896in;
%original-height 3.1249in;  cropleft "0";  croptop "1";  cropright "1";
%cropbottom "0";  filename 'LxyBCsAu.eps';file-properties "XNPEU";}}}%
%BeginExpansion
{\includegraphics[
height=2.7854in,
width=3.6364in
]%
{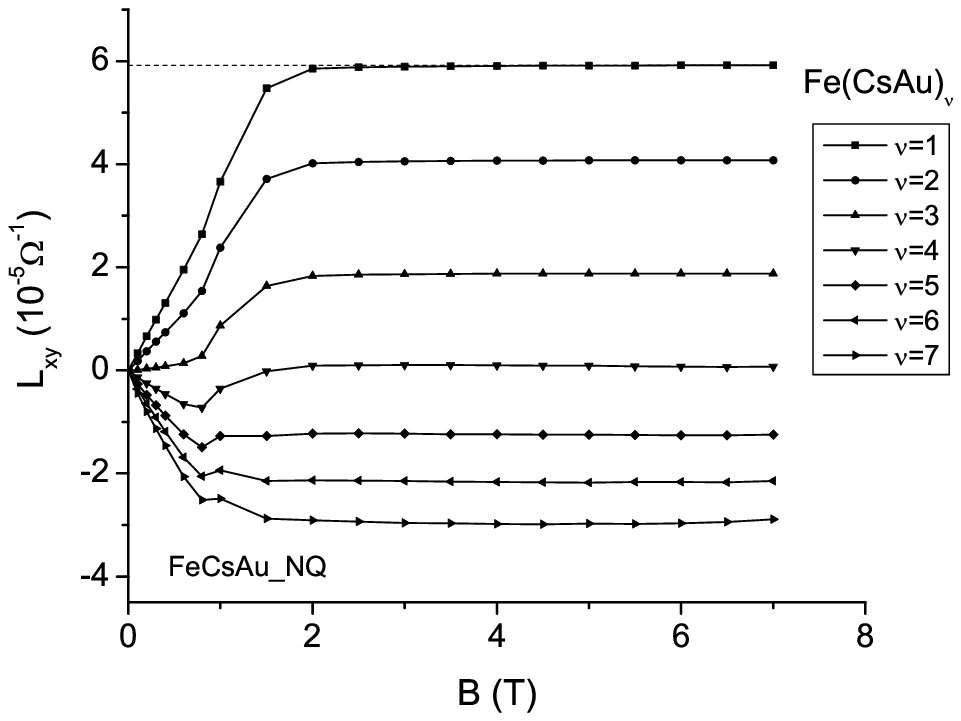}%
}%
%EndExpansion
\\
&
\begin{tabular}
[c]{l}%
Fig.4: The anomalous Hall conductance of an Fe(CsAu)$_{\nu}$ multi\\
layer as a function of the applied magnetic field for different\\
numbers of layers $\nu$.
\end{tabular}
\end{align*}

In Fig.5 the zero field AHC $L_{xy}^{0}$ is plotted versus the thickness of
the multi-layers of (CsAu)$_{\nu}$ and (CsAg)$_{\nu}$ (the Fe thickness
subtracted). Obviously the two systems behave very differently.%

\begin{align*}
&
%TCIMACRO{\FRAME{itbpF}{3.1307in}{2.6251in}{0in}{}{}{lxycsagau.eps}%
%{\special{ language "Scientific Word";  type "GRAPHIC";
%maintain-aspect-ratio TRUE;  display "USEDEF";  valid_file "F";
%width 3.1307in;  height 2.6251in;  depth 0in;  original-width 3.7825in;
%original-height 3.1664in;  cropleft "0";  croptop "1";  cropright "1";
%cropbottom "0";  filename 'LxyCsAgAu.eps';file-properties "XNPEU";}}}%
%BeginExpansion
{\includegraphics[
height=2.6251in,
width=3.1307in
]%
{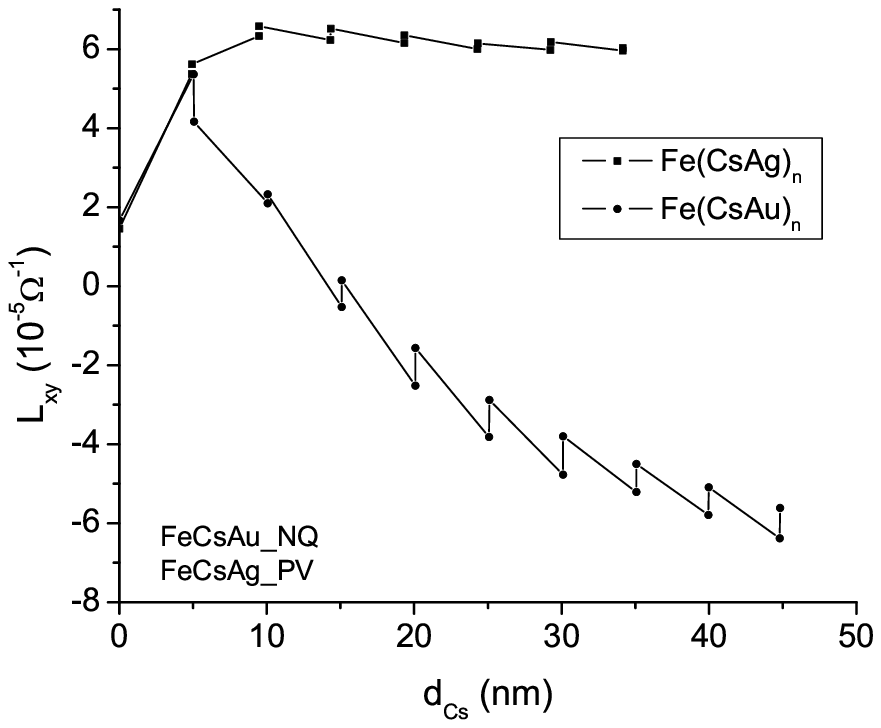}%
}%
%EndExpansion
\\
&
\begin{tabular}
[c]{l}%
Fig.5: The extrapolated anomalous Hall conductance for an\\
Fe(CsAg)$_{\nu}$ and the Fe(CsAu)$_{\nu}$ multi-layer for successive\\
condensation of the Cs and noble metal layers. The abscissa\\
is the total Cs thickness $d_{Cs}$ in the multi-layers.
\end{tabular}
\end{align*}%
\[
\]

\section{Discussion}

\subsection{The Fe(CsAg)$_{\nu}$ multi-layers}

The pure but very disordered Fe film has an AH conductance of about
$1.5\times10^{-5}\Omega^{-1}$. When the first Cs film is superimposed the AH
conductance increases to a value of about $5.5\times10^{-5}\Omega^{-1}$. This
induced AHE is a well known phenomena which has been analyzed by one of the
authors \cite{B105}, \cite{B106} in a simple model. Its origin is briefly
recalled in the appendix.

\subsection{The Fe(CsAu)$_{\nu}$ multi-layers}

Since the mean free path in Fe(CsAu)$_{\nu}$ is essentially identical to that
in the Fe(CsAg)$_{\nu}$ multi-layers it experiences the same induced AH
conductance as the Fe(CsAg)$_{\nu}$ multi-layers. But there is obviously an
additional effect. In this paper we will focus on the additional effect, i.e.,
the difference between the two multi-layers.

The AHC drops clearly when the first 0.04 atomic layer of Au are condensed
onto the Cs. The difference between the CsAu and the CsAg has to be due to the
spin-orbit scattering of the Au.\ It indicates that there is a spin current in
the Cs layer.

To simplify the discussion we have plotted in Fig.6 the difference in the AHC
between the Fe(CsAg)$_{\nu}$ and the Fe(CsAu)$_{\nu}$ multi-layers. It is
important to remember that the additional contribution of the Fe(CsAu)$_{\nu}$
is negative and Fig.6 shows the absolute value of the difference. This AHC
$\Delta L_{xy}^{0}$ is caused by the spin-orbit scattering of the Au. Again
this proves that we have a spin current in the Cs film. The Au atoms measure
the local spin current density, and $\Delta L_{xy}^{0}$ is proportional to the
integrated spin current density.%

\begin{align*}
&
%TCIMACRO{\FRAME{itbpF}{3.3001in}{2.6857in}{0in}{}{}{lxyagau.eps}%
%{\special{ language "Scientific Word";  type "GRAPHIC";
%maintain-aspect-ratio TRUE;  display "USEDEF";  valid_file "F";
%width 3.3001in;  height 2.6857in;  depth 0in;  original-width 3.8638in;
%original-height 3.1399in;  cropleft "0";  croptop "1";  cropright "1";
%cropbottom "0";  filename 'LxyAgAu.eps';file-properties "XNPEU";}}}%
%BeginExpansion
{\includegraphics[
height=2.6857in,
width=3.3001in
]%
{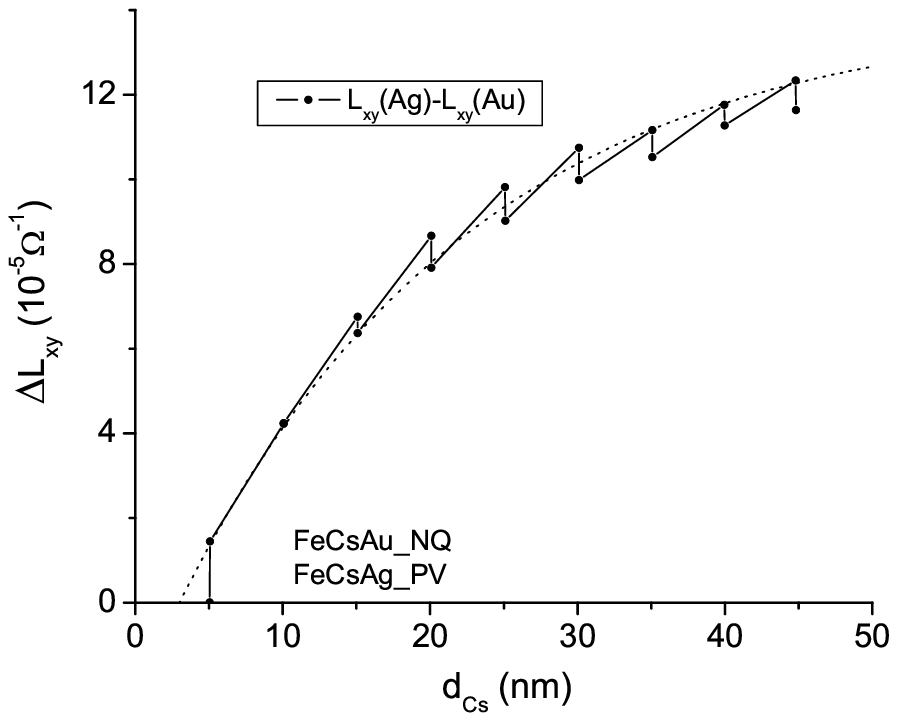}%
}%
%EndExpansion
\\
&
\begin{tabular}
[c]{l}%
Fig.6: The (negative) spin Hall effect in Fe(CsAu)$_{\nu}$ multi-layer.\\
It is the difference in the AHC between the Fe(CsAg)$_{\nu}$ and the\\
Fe(CsAu)$_{\nu}$ multi-layers. The dotted curve represents an exponential\\
fit with $\Delta L_{xy}=1.4\times10^{-4}$ $\ast\left\{  1-\exp\left[
-\frac{\left(  d_{CsAu}-3nm\right)  }{20nm}\right]  \right\}  \Omega^{-1}$%
\end{tabular}
\end{align*}

First we discuss the effect of the combined CsAu evaporations. The magnitude
of $\Delta L_{xy}^{0}$ increases with $\nu$ and levels off for larger
thicknesses of the (CsAu)$_{\nu}$. This means that the spin current extends
quite deep into the (CsAu)$_{\nu}$ multi-layer. We expect that the electrons
in the (CsAu)$_{\nu}$ are essentially unpolarized except very close to the
FeCs interface. Therefore the spin current is solely caused by the different
drift velocities of the spin up and down electrons. As a consequence a
scattering process that destroys the drift velocity will also destroy the spin
current. Therefore we expect that the spin current decays exponentially with
the distance from the interface as $j_{s}^{0}\exp\left(  -z/l_{chr}\right)  $
and the decay length $l_{chr}$ is of the order of the MFP. If the Au atoms
would be homogeneously distributed in the Cs then the AH conductance would
integrate the spin Hall effect over the film thickness. This yields for
$\Delta L_{xy}$%
\[
\Delta L_{xy}\varpropto\int_{0}^{d}\exp\left(  -z/l_{chr}\right)
dz\varpropto\left[  1-\exp\left(  -\frac{d}{l_{chr}}\right)  \right]
\]

In reality the Au impurities are located at discrete distances of $\nu d_{Cs}$
($\nu=1,2..$) from the FeCs interface. This modifies the result for $\Delta
L_{xy}$ slightly. We find a good fit of the experimental data with the
function $\Delta L_{xy}=$ $1.4\times10^{-4}$ $\ast\left\{  1-\exp\left[
-\left(  d_{CsAu}-3nm\right)  /20nm\right]  \right\}  \Omega^{-1}$. This
function is plotted in Fig.6 as a dashed curve. The experimental
characteristic length is $l_{chr}=20nm$. In the appendix we derive the spatial
dependence of the spin current.

\subsubsection{The Au-surface effect}

Next we discuss the effect of the Cs and Au evaporation in more detail. Only
for $\nu=1$ does the Au condensation yield an increase of the AHC. For $\nu=2$
the effect of the Au condensation is very small and for larger $\nu$ the AHC
actually decreases. From Fig.2 one recognizes that the condensation of Au onto
the Cs reduces the conductance of the multi-layer. This reduction is
equivalent to the conductance of about $2nm$ to $2.5nm$ of Cs.

This agrees well with a rather surprizing observation that we made in a
previous investigation. There we condensed Pb and Au onto an FeCs double
layer. The Pb and Au introduced an additional AHC, but the dependence of the
$\Delta L_{xy}$ on the coverage of the surface impurities was quite
unexpected. For a coverage of $d_{Pb}$ or $d_{Au}$ of about 0.03 atomic layer
$\Delta L_{xy}$ showed a large extremum, and at a coverage of 0.2 atomic layer
$\Delta L_{xy}$ had almost completely disappeared. Our interpretation of this
behavior was that impurities at the surface introduced so much disorder close
to the surface that (i) most of the spin current decayed before\ it reached
the Pb (Au) impurities at the surface and (ii) the spin-orbit scattered
electrons could not generate much current in y-direction because their MFP was
so short. Our new experimental data in Fig.5 or Fig.6 demonstrate that this
reduction of the MFP is so dominant that after each Au evaporation $\Delta
L_{xy}$ reduces. The new Au impurities do not generate any significant
additional AHC, and the AHC of the lower Au layer is reduced because of the
MFP reduction.

This reduction of the MFP in the last Cs layer can be (i) a reduction in the
whole last Cs layer by a factor of two or (ii) a reduction to zero in the
upper $2nm$ of Cs. The latter case corresponds to electron localization at the
upper surface due to the impurities. In the appendix we will give some
arguments for the latter case.%

\[
\]

\section{Conclusion}

The anomalous Hall effect in Fe(CsAu)$_{\nu}$ multi-layers has been
investigated. It is compared with the AHE of Fe(CsAg)$_{\nu}$ multi-layers.
Although the two kinds of impurities cause almost identical normal scattering
and the MFPs in the resulting multi-layers are essentially the same, the AHE
is quite different. The difference in the AH conductance $\Delta L_{xy}$ is
due to a spin current which experiences the spin-orbit scattering of the Au
impurities. This effect has been baptized "Spin Hall Effect". The spin current
is due to the different specular reflection of spin up and down electrons at
the FeCs interface. By condensing a series of CsAu layers the size of the spin
current can be analyzed geometrically. It decays as $\left[  1-\exp\left(
-z/l_{chr}\right)  \right]  $ where $z$ is the distance from the interface and
$l_{chr}$ is a length which has a value of $l_{chr}=20nm$.

During the preparation of the multi-layers a thin 0.04 atomic layer thick
Au\ layer is condensed onto a fresh $5nm$ thick Cs layer. Each time the
longitudinal conductance $L_{xx}$ is reduced by $\Delta L_{xx}.$This $\Delta
L_{xx}$ corresponds to the conductance of about $2nm$ of Cs. It is suggested
that this is a (precursor of) surface localization of the conduction electrons
in the Cs. Such an effect has recently been observed in the superconducting
proximity effect of PbK double layers \cite{B134}.

\newpage

\section{Appendix}

\subsection{Origin of the induced anomalous Hall effect}

The pure but very disordered Fe film in the Fe(CsAg)$_{\nu}$ multi-layer has
an AH conductance of about $1.5\times10^{-5}\Omega^{-1}$. When Fe is covered
with the CsAg film the AH conductance increases with the coverage (as is shown
in Fig.5). This is due to the following facts \cite{B105}, \cite{B106}:

(i) If the electric field in the Fe points in the x-direction then the current
has a finite y-component (which represents the AH conductance of the Fe film
$L_{yx}($Fe$)=d_{Fe}\ast j_{y}/E_{x}$). A fraction of the conduction electrons
cross the interface to the Cs and carry with them their y-component of the
current. This current in y-direction yields an additional contribution the
Hall conductance $L_{xy}$. This induced AH conductance increases with the Cs
thickness. It saturates when the Cs thickness reaches about half the mean free
path in the Cs.

(ii) The second contribution is caused by the electrons which are accelerated
in the Cs. When those electrons cross into the Fe they have a much larger
drift velocity than the native Fe electrons. They therefore yield a strongly
enhanced AH conductance within a thin layer of the Fe. This layer has a
thickness of about half the mean free path of the Fe electrons.

Both contributions are of the same order of magnitude and contribute to the
increase in the AH conductance of the Fe(CsAg)$_{\nu}$ multi-layer. The role
of the Ag impurities lies in the limitation of the mean free path to about
$17nm$. Therefore the increase of the AH conductance levels out at about
$10nm$. This mechanism of the induced AHE depends on AHE in the Fe layer and
the mean free paths in the Fe and the Cs.

\subsection{Origin of the spin current}

We can imagine two different origins of the spin current in the Cs film, (i)
transfer of the spin current in the ferro magnetic Fe film through the FeCs
interface and (ii) different specular reflection of the spin up and down
electrons in the Cs at the FeCs interface. The experimental data favor the
second mechanism as the following discussion shows.

(i) The spin up and down electrons in the Fe have different densities and
MFP's. Therefore the spin up and down current densities in the Fe are
different. This represents a spin current which could cross the FeCs interface
and enter the Cs film. The resulting spin current density $j_{s}$ at the upper
Cs surface should be at best independent of the Cs thickness. Any scattering
within the Cs should reduce $j_{s}$ at the free Cs surface. Experimentally we
observe, however, a larger AH conductance for the first Au coverage when the
Cs thickness is larger. In a previous experiment we covered a FeCs double
layer ($d_{cs}=27.5nm$) with 0.02 atomic layer of Au and observed $\Delta
L_{xy}=$ $-8.8\ast10^{-5}\Omega^{-1}$. In the present experiment we covered
the FeCs double layer ($d_{cs}=5nm$) with 0.04 atomic layer of Au and observed
$\Delta l_{xy}=1.2\times10^{-5}\Omega^{-1}$. Although the Au coverages are not
identical it is obvious that the larger Cs thickness yields a large spin
current density at the free surface. \newline(ii) The spin up and down
electrons in the Cs experience a different degree of specular reflection at
the FeCs interface. These electrons have accumulated a finite drift velocity
before they are reflected. This drift velocity increases with larger Cs
thickness because the effective MFP increases. The resulting spin current
after the reflection is proportional to the current before the reflection.
Therefore the spin current density at the free surface increases with
increasing Cs thickness. This is the experimental observation.

\subsection{Theoretical description of the spin current}

\subsubsection{Homogeneous spin current}

Let us first consider a spin current in a non-magnetic metal film with a
constant spin current density $j_{s}$. The metal contains normal and
spin-orbit scattering impurities. The total relaxation time $\tau_{0}$ is the
same for spin up and down electrons. In addition the SOS impurities scatter
spin up and down electrons to the opposite film edges. One of the authors
calculated the AH conductance for a spin current in the presence of SOS
\cite{B126}. It can be expressed by an anomalous Hall cross section $a_{xy}$.
(We use the symbol $a$ for the cross section instead of $\sigma$ to avoid
confusion with the conductivity $\sigma$). If we restrict ourselves to
(s,p)-scattering then the dominant term of the cross section $a_{xy}$ is
\begin{equation}
a_{xy}\thickapprox-\frac{4\pi}{3k_{F}^{2}}\sin\left(  \delta_{1,+}%
-\delta_{1,-}\right)  \sin\delta_{0}\cos\left(  \delta_{1,+}+\delta
_{1,-}-\delta_{0}\right)  \label{axy}%
\end{equation}
(The full, lengthy expression for $a_{xy}$ is given in ref. \cite{B126}). Here
$k_{F}$ is the Fermi wave vector and $\delta_{l,\pm}=\delta_{l\pm1/2,l}$ are
the Friedel phase shifts for spin up and down electrons with the orbital and
total angular momenta $l$ and $j=l\pm1/2$. The physical meaning of this cross
section is the following: A current density $j_{x,\uparrow}$ of spin up
electrons ($\uparrow\parallel\widehat{\mathbf{z}}$) encounters a SOS impurity.
Due to the asymmetric SOS a fraction of the current is deflected in
y-direction. This fraction is a current $I_{y,\uparrow}$ which is equal to the
current density times the cross section $\alpha_{xy}$: $I_{y,\uparrow}%
=a_{xy}j_{x,\uparrow}$. This current decays after the time $\tau_{0}$ i.e.,
after the distance of the MFP $l_{0}$. For a SOS impurity concentration
$n_{i}$ one obtains an AH current density in y-direction of%
\[
j_{y,\uparrow}=l_{0}n_{i}a_{xy}j_{x,\uparrow}%
\]
For the spin down electrons one obtains an AH current density with the
opposite sign%
\[
j_{y,\downarrow}=-l_{0}n_{i}a_{xy}j_{x,\downarrow}%
\]
So the total current density in the y-direction is
\begin{align*}
j_{y}  &  =l_{0}n_{i}a_{xy}\left(  j_{x,\uparrow}-j_{x,\downarrow}\right)
=pl_{0}n_{i}a_{xy}j_{c}\\
&  =pn_{i}a_{xy}\frac{ne^{2}}{\hbar k_{F}}l_{0}^{2}E_{x}%
\end{align*}
where $p=\left(  j_{x\uparrow}-j_{x\downarrow}\right)  /\left(  j_{x\uparrow
}+j_{x\downarrow}\right)  $ is the polarization of the current density. The
resulting (spin Hall) conductivity is
\begin{equation}
\sigma_{xy}=pn_{i}a_{xy}\frac{ne^{2}}{\hbar k_{F}}l_{0}^{2} \label{sig_xy0}%
\end{equation}

The resulting AH angle is $\tan\alpha=j_{y}/j_{x}=pl_{0}n_{i}a_{xy}$. The AH
angle, the MFP and the impurity density are experimentally known. The AH cross
section $a_{xy}$ is known in terms of the Friedel phase shifts. If someone
would calculate the Friedel phase shifts $\delta_{l\pm1/2,l}$ one could
directly measure the polarization of the current by means of the AH conductance.

\subsubsection{Inhomogeneous spin current}

In our experiment the FeCs interface acts the source of spin current. The
appropriate treatment would be using the Boltzmann equation. If one includes
the size effect in the thin films then the problem becomes rather involved. We
treat the problem here in a simplified fashion. We divide the charge current
density $j_{c}\left(  z\right)  $ in the Cs film into $\left[  j^{+}\left(
z\right)  +j^{-}\left(  z\right)  \right]  .$ The superscript gives the sign
of the $k_{z}$-component of the involved electrons. Here $j_{c}^{-}\left(
z\right)  =$ $\left[  j_{\uparrow}^{-}\left(  z\right)  +j_{\downarrow}%
^{-}\left(  z\right)  \right]  $ is the part of the charge current with
$k_{z}<0$ which moves towards the interface. Similarly $j_{s}^{+}\left(
z\right)  =\left[  j_{\uparrow}^{+}\left(  z\right)  -j_{\downarrow}%
^{+}\left(  z\right)  \right]  $ is the (part of the) spin current with
$k_{z}>0$ which moves away from the surface. At the FeCs interface the spin up
and down electrons experience a different degree of specular reflection,
$\lambda_{\uparrow}$ and $\lambda_{\downarrow}$. The specular part of the
reflected electrons maintains the accummulated drift velocity and contributes
a current density $j^{+}\left(  0\right)  =\lambda_{\uparrow}j_{\uparrow}%
^{-}\left(  0\right)  +$ $\lambda_{\downarrow}j_{\downarrow}^{-}\left(
0\right)  $ at $z=0$. When $\lambda_{\uparrow}\neq\lambda_{\downarrow}$ then
the interface generates a spin current $j_{s}^{+}$ which moves away from the
interface into the Cs and is given by $j_{s}^{+}\left(  0\right)  =\lambda
j_{c}^{-}\left(  0\right)  $ where $\lambda=\lambda_{\uparrow}-\lambda
_{\downarrow}$. \ From now on the electric field has no effect on the size of
the spin current.

The electrons (of the spin current) with $k_{z}>0$ move with an average
velocity $\alpha v_{F}$ away from the interface into the Cs. Here $v_{F}$ is
the Fermi velocity and $\alpha$ is of the order of $1/2.$ Within the Cs film
both electron spins experience the same relaxation time $\tau_{0}$. Therefore
$\tau_{0}$ is the decay time for both the charge current and the spin current,
for example $dj_{s}^{+}/dt=-j_{s}^{+}/\tau_{0}$. In the stationary situation
this yields a spatial decay of the spin current.
\[
\frac{\partial j_{s}^{+}}{dz}=\frac{dj_{s}^{+}}{dt}/\frac{dz}{dt}=-\frac
{1}{\tau_{0}}\frac{1}{\alpha v_{F}}j_{s}^{+}%
\]
which yields
\begin{align}
j_{s}^{+}\left(  z\right)   &  =j_{s}^{+}\left(  0\right)  \exp\left(
-\frac{z}{\alpha v_{f}\tau_{0}}\right)  =j_{s}^{+}\left(  0\right)
\exp\left(  -\frac{z}{l_{chr}}\right) \label{s_c+}\\
&  =\lambda j_{c}^{-}\left(  0\right)  \exp\left(  -\frac{z}{l_{chr}}\right)
=\frac{\lambda}{2}j_{c}\left(  0\right)  \exp\left(  -\frac{z}{l_{chr}}\right)
\nonumber
\end{align}
If there is a finite specular reflection $r$ at the upper surface then part of
the spin current $j_{s}^{+}$ is reflected at the upper surface. This results
in a spin current which propagates in the negative z-direction and has the
form
\[
j_{s}^{-}\left(  z\right)  =j_{s}^{-}\left(  d_{Cs}\right)  \exp\left(
-\frac{d_{Cs}-z}{l_{chr}}\right)  =\frac{r\lambda}{2}j_{c}\left(  0\right)
\exp\left(  -\frac{2d_{Cs}-z}{l_{chr}}\right)
\]

In the following we neglect the specular reflection at the upper surface and
we treat the charge current density as constant in the film. Then the
polarization $p\left(  z\right)  $ of the current density is position
dependent.%
\[
p\left(  z\right)  =\frac{j_{\uparrow}\left(  z\right)  -j_{\downarrow}\left(
z\right)  }{j_{\uparrow}\left(  z\right)  +j_{\downarrow}\left(  z\right)
}=\frac{\lambda}{2}\exp\left(  -\frac{z}{l_{chr}}\right)
\]

To obtain the total AH conductance $L_{xy}$ one has to integrate equation
(\ref{sig_xy0}) over $dz.$%
\begin{align}
L_{xy}  &  =n_{i}a_{xy}\frac{ne^{2}}{\hbar k_{F}}l_{0}^{2}\int_{0}^{d}p\left(
z\right)  dz\nonumber\\
&  =\frac{1}{2}\lambda n_{i}a_{xy}\frac{ne^{2}}{\hbar k_{F}}l_{0}^{2}%
l_{chr}\left[  1-\exp\left(  -\frac{d_{Cs}}{l_{chr}}\right)  \right]
\label{Lxy}%
\end{align}

Our experimental results agree within the accuracy of the measurement with the
contribution of equation (\ref{Lxy}).

\subsection{Interface localization}

Our group recently investigated the effect of surface impurities on the
conductance of alkali films using the superconducting proximity effect
\cite{B134}. In these experiments it was demonstrated that the conductance of
a $5nm$ thick K film on top of a Pb film lost its conductance when covered
with a sub-mono layer of Pb. But it maintained its full ability to reduce the
superconducting transition temperature of the Pb film. As discussed in
\cite{B134} this is the behavior of localized electrons in the K whose
localization length is larger than the film thickness. Then electrons in the
layer can move perpendicular but not parallel to the layer.

Such a surface localization would explain why the AHE of Pb or Au impurities
on top of an FeCs double layer disappears when the impurity coverage exceeds
0.03 atomic layer. The conductance and AHC plot in Fig.2 and Fig.5,6 suggest
that the Au coverage repeatedly introduces surface localization or its
precursor for the Cs conduction electrons. The condensation of the next Cs
layer essentially removes this localization so that the Au impurities
contribute to the AHC. It is also likely that the MFP of the conduction
electrons in the (CsAu)$_{\nu}$ layers is anisotropic. The perpendicular MFP
will be larger than the parallel one.

\subsection{Local mean free path}

In Fig.3 the average MFP of the conduction electrons in Cs with Au or Ag
impurities is plotted. In this evaluation the conductance of the whole Cs film
divided by the total Cs is evaluated. The MFP is calculated from the resulting
conductivity. This evaluation is appropriate for pure films. The structure of
their initial layer recrystallizes somewhat when the initial layer is covered
with the same material. When, however, the individual Cs layers are separated
by a thin Au or Ag sub-mono layer then the structure of the previous layers
can be frozen. In this case a local MFP might be more appropriate. It is
calculated from the local conductivity $\Delta L_{xx}/5nm.$ $\Delta L_{xx}$ is
the conductance of the multi-layer minus the conductance of the last
Fe(CsAu)$_{\nu}$. For the upper points in Fig.6 one has $\Delta L_{xx}%
=L_{xx}\left(  \text{Fe(CsAu)}_{\nu}\text{Cs}\right)  -L_{xx}\left(
\text{Fe(CsAu)}_{\nu}\right)  $ and for the lower points $\Delta L_{xx}%
=L_{xx}\left(  \text{Fe(CsAu)}_{\nu+1}\right)  -L_{xx}\left(  \text{Fe(CsAu)}%
_{\nu}\right)  $
\begin{align*}
&
%TCIMACRO{\FRAME{itbpF}{3.0344in}{2.5255in}{0in}{}{}{locmfpcsau.eps}%
%{\special{ language "Scientific Word";  type "GRAPHIC";
%maintain-aspect-ratio TRUE;  display "USEDEF";  valid_file "F";
%width 3.0344in;  height 2.5255in;  depth 0in;  original-width 3.8082in;
%original-height 3.1664in;  cropleft "0";  croptop "1";  cropright "1";
%cropbottom "0";  filename 'locMfpCsAu.eps';file-properties "XNPEU";}}}%
%BeginExpansion
{\includegraphics[
height=2.5255in,
width=3.0344in
]%
{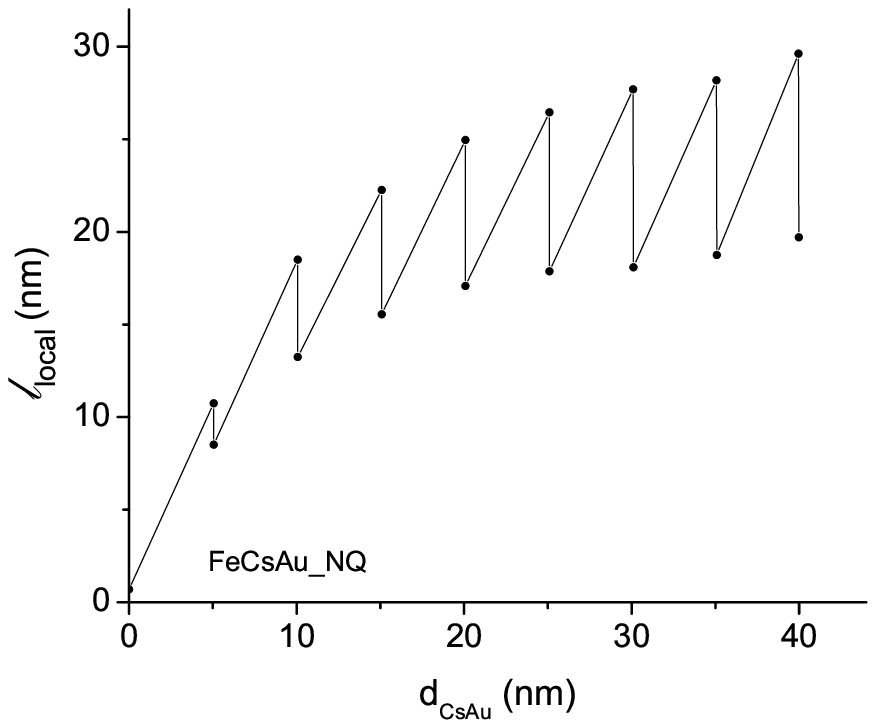}%
}%
%EndExpansion
\\
&
\begin{tabular}
[c]{l}%
Fig.6: The local mean free path in the Fe(CsAu)$_{\nu}$\\
multi-layer.
\end{tabular}
\end{align*}

\subsection{The normal Hall effect in Fe(CsAu)$_{\nu}$ multi-layers}

The measurement of the Hall effect of the multi-layers also yields the Hall
constant. Fig.7 shows the local Hall constant as a function of the Cs
thickness. The upper points represent the last Cs mini layer and the lower
points the last Cs layer with Au coverage. While the Hall constant of the
individual Cs layers lies in the range of $-\left(  63\pm3\right)
\times10^{-11}m^{3}/As$ the condensation of the noble metal increases the
value to $-\left(  77\pm2\right)  \times10^{-11}m^{3}/As$. The condensation of
the Au or Ag impurities acts as if 25\% of the Cs layer do no longer
contribute to the normal Hall effect.
\begin{align*}
&
%TCIMACRO{\FRAME{itbpF}{3.2644in}{2.7015in}{0in}{}{}{chcsau.eps}%
%{\special{ language "Scientific Word";  type "GRAPHIC";
%maintain-aspect-ratio TRUE;  display "USEDEF";  valid_file "F";
%width 3.2644in;  height 2.7015in;  depth 0in;  original-width 3.7825in;
%original-height 3.1249in;  cropleft "0";  croptop "1";  cropright "1";
%cropbottom "0";  filename 'ChCsAu.eps';file-properties "XNPEU";}}}%
%BeginExpansion
{\includegraphics[
height=2.7015in,
width=3.2644in
]%
{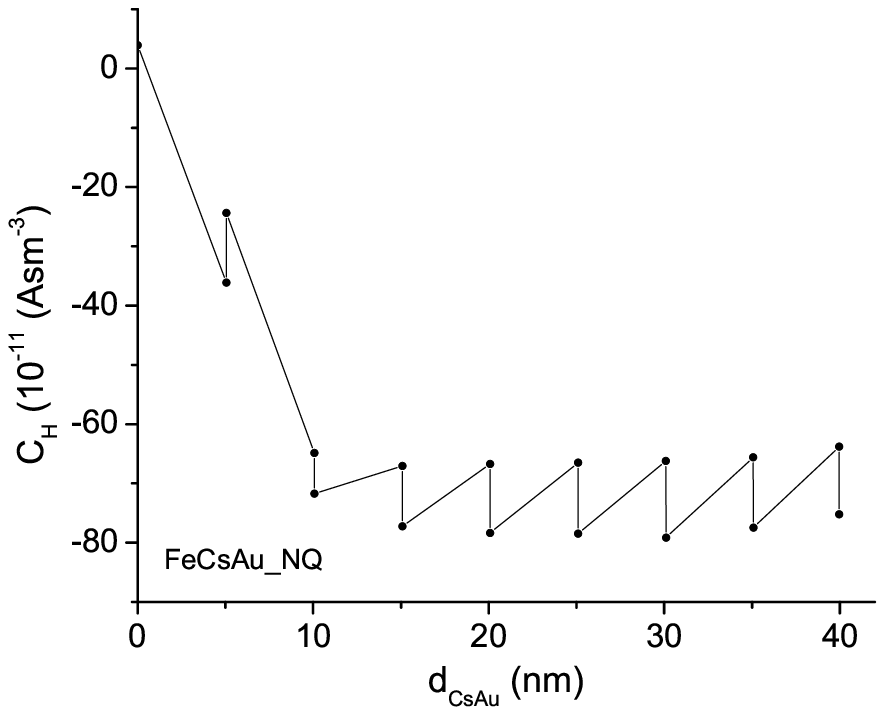}%
}%
%EndExpansion
\\
&
\begin{tabular}
[c]{l}%
Fig.7: The local Hall constant for the $\text{Fe(CsAu)}_{\nu}$ multi-layers\\
as function of the total Cs thickness.
\end{tabular}
\end{align*}

\subsection{Asymmetric Hall curve}

In Fig.8 the Hall curves for Fe(CsAu)$_{4}$ and Fe(CsAg)$_{4}$ are plotted as
a function of the magnetic field. The normal Hall conductance is subtracted,
but the curves are not symmetrized. It is very obvious that the Hall curve for
Fe(CsAu)$_{4}$ is not anti-symmetric. Such a behavior can be easily observed
in the magnetoresistance of a ferro magnetic film or multi-layer. It occurs
when the magnetic field is aligned perfectly perpendicular to the film. If the
magnetic field is lowered and approaches $B=0$ then the magnetic domains align
parallel to the film plane. The orientation within the film plane is
arbitrary. Therefore the angle between the current and the magnetization of
the single domain will be arbitrary too. Generally the resistance is
anisotropic and depends on the angle between current and magnetization. Then
the resistance at zero and small fields depends on the accidental orientation
of the magnetic domains within the film plane. One observes hysteresis. This
can be avoided by slightly tilting the film substrate so that the magnetic
field is no longer perfectly perpendicular to the film.

On the other hand such hysteresis is generally not observed for the AHE. Only
the z-component of the magnetization $M$ contributes to the AHE. The
z-component of the magnetization, $M_{z}$, is equal to the applied magnetic
flux $B$ (for $B<M$). This is demonstrated by the perfectly anti symmetric
shape of the AH curve of Fe(CsAg)$_{4}$. Therefore we believe that the
asymmetric shape of the AH curve for Fe(CsAu)$_{4}$ contains interesting
information about the system. It might suggest that the orientation of the
electron spins in this field range is not parallel to the magnetic field. So
far we have not investigated further this observation. It does not effect any
of our conclusions because we only used the AHC for the large fields and
extrapolated the linear part back to zero.%
\begin{align*}
&
%TCIMACRO{\FRAME{itbpF}{3.56in}{2.9058in}{0in}{}{}{lxy4.eps}%
%{\special{ language "Scientific Word";  type "GRAPHIC";
%maintain-aspect-ratio TRUE;  display "USEDEF";  valid_file "F";
%width 3.56in;  height 2.9058in;  depth 0in;  original-width 3.77in;
%original-height 3.0726in;  cropleft "0";  croptop "1";  cropright "1";
%cropbottom "0";  filename 'Lxy4.eps';file-properties "XNPEU";}}}%
%BeginExpansion
{\includegraphics[
height=2.9058in,
width=3.56in
]%
{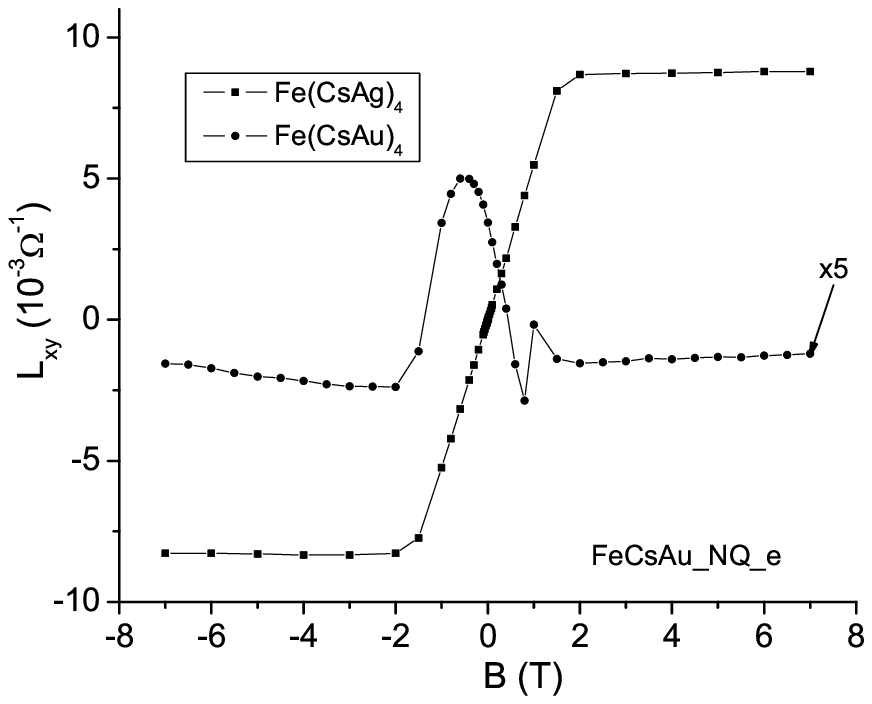}%
}%
%EndExpansion
\\
&
\begin{tabular}
[c]{l}%
Fig.8: The anomalous Hall curves for $\text{Fe(CsAu)}_{4}$ and
$\text{Fe(CsAg)}_{4}$\\
as a function of the magnetic field. The normal Hall effect is\\
subtracted but the curves are not symmetrized.
\end{tabular}
\end{align*}

Acknowledgement: The research was supported by the National Science Foundation
NIRT program, DMR-0439810.

\end{document}